\documentclass[aps,prl,showpacs,twocolumn,superscriptaddress]{revtex4}
\usepackage{graphicx}
\usepackage{bm}
\usepackage{amsmath}

\def\be{\begin{equation}}
\def\ee{\end{equation}}
\def\bea{\begin{eqnarray}}
\def\eea{\end{eqnarray}}

\begin{document}
\title{Quantum Annealing Effect on Entropic Slowing Down in Frustrated Decorated Bond System}

\author{Shu Tanaka}
\email[Electronic address: ]{shu-t@spin.phys.s.u-tokyo.ac.jp} 
\affiliation{Department of Physics, The University of Tokyo,
7-3-1, Hongo, Bunkyo-ku, Tokyo 113-0033, Japan}

\author{Seiji Miyashita}
\email{miya@spin.phys.s.u-tokyo.ac.jp}
\affiliation{Department of Physics, The University of Tokyo,
7-3-1, Hongo, Bunkyo-ku, Tokyo 113-0033, Japan}
\affiliation{CREST, JST, 4-1-8 Honcho Kawaguchi, Saitama, 332-0012, Japan}

\begin{abstract}
We propose that the importance of the quantum annealing procedure to find the ground state of 
frustrated decorated bond systems where '{\it entropic slowing down}' happens due to peculiar density of states. 
Here, we use the time dependent Schr\"odinger equation to analyze the real time dynamics of the process.
It is found that the quantum annealing is very efficient comparing to the thermal annealing for 
searching the ground state of the systems.
We analyze the mechanism of quantum annealing from a view point of adiabatic process.
\end{abstract}

\maketitle

\section{Introduction}\label{}
Optimization problem is an important topic in vast area of science. 
The most typical example is searching the ground state of random spin system such as the spin glass. 
The free energy landscape of random spin system is very complex. 
Therefore, we encounter the difficulty that the system does not reach 
the ground state but stays in the metastable one. 
There are two types of slowing down. One is the case where the state is 
trapped at a energetically metastable state. 
An energy barrier prevents the state from escaping to the equilibrium state.
This is '{\it energetic slowing down}'.
On the other hand, though there is no energy barrier, 
the system can not reach the equilibrium state in a short time which has been found 
in the regularly decorated bond system \cite{Tanaka1},\cite{Tanaka2}.
We called this situation '{\it entropic slowing down}'.
In the latter case, we have found that effective time scale $\tau_0$ of the ordering process 
becomes extremely long and the relaxation process is practically frozen.
In this case it is difficult to find the energy minimam stracture using the thermal annealing process.

Kadowaki and Nishimori \cite{Kadowaki} have proposed
the quantum annealing method for the energetic slowing down system.
Afterward, the efficiency of the quantum annealing method has been confirmed
in the energetic slowing down system \cite{Santoro},\cite{Suzuki}.
Our aim of this study is to demonstrate the efficiency of the quantum annealing method 
in the entropic slowing down system.

\section{Model}
We study the decorated bond system depicted in Fig. 1 as an example of systems with
"entropically slowing down". 
The Hamiltonian of this system is
\[
\mathcal{H}_0 = -J' \sigma_1^z \sigma_2^z -J \sigma_1^z \sum_{i=1}^N S_i^z 
-J \sigma_2^z \left( \sum_{i=1}^{\frac{N}{2}} S_i^z - \sum_{i=\frac{N}{2}+1}^N S_i^z\right),
\]
where $N$ is the number of the decorated spin $\left\{ S_i \right\}$.
The effective coupling $J_{\mathrm{eff}}$
between $\sigma_1^z$ and $\sigma_2^z$ is defined 
$\left\langle \sigma_1^z \sigma_2^z \right\rangle = \tanh \beta J_{\mathrm{eff}}$.
In the case of Fig. 1, the contributions through the decoration spins are canceled out 
each other and $J_{\mathrm{eff}} = J'$ due to the direct interaction. 
However, the dynamics is not simple as the case of the single bond.
If the system is trapped at the state depicted in Fig. 1-(b),
it does not easily reach the equilibrium state due to the entropy effect \cite{Tanaka1}. 
In this case, the thermal annealing is not efficient and the relaxation time is about
$2^{N/2}$ for this system to reach the ground state \cite{Tanaka2}.
Here, we introduce quantum annealing using the time dependent transverse field 
$\mathcal{H}_t\left( t \right) = -\Gamma \left( t \right) \left( \sigma_1^x + \sigma_2^x + \sum_i S_i^x \right)$, 
and we change the transverse field as $\Gamma \left( t\right)=\Gamma_0 \left(1-t/\tau \right)$.

%%%%%%%%%%%%%%%%%%%% Kondo in title, abstract and/or keywords %%%%%%%%%%%%%%%%
  \begin{figure}[h]
  \begin{center}
\includegraphics[scale=1]{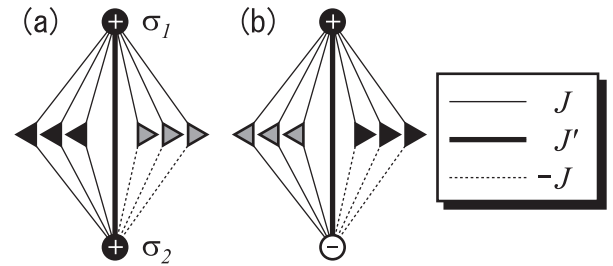}
\end{center}
\caption{ 
The thick and thin lines denote the ferromagnetic coupling $J'>0$ and $J>0$, respectively.
The dotted lines denote the antiferromagnetic coupling $-J<0$. 
The black triangles denote $+$ spin and the gray spins are not fixed ($+$ or $-$) spin.
(a) The ground state with the correlation function $\sigma_1^z \sigma_2^z = +1$.
(b) The '{\it entropically metastable state}' with $\sigma_1^z \sigma_2^z = -1$. 
In order to relax to (a) from this configuration,
all the gray spins must be align to cancel the interaction from the black spins.
But the probability of happening of this situation is very small. 
This is entropic slowing down.
}
  	\label{fig-1}
  \end{figure}
%%%%%%%%%%%%%%%%%%%%%%%%%%%%%%%%%%%%%%%%%%%%%%%%%%%%%%%%%%%%%%%%%%%%%%%%%%%%%%

\section{Real Time Dynamics of the Decorated Bond System by Schr\"odinger Equation}
We study the real time dynamics of the decorated bond system with the time dependent transverse field.
Here, the time evolution of the state vector $\left| \Phi \left( t \right) \right\rangle$ is 
determined by the Schr\"odinger equation.
The initial state $\left| \Phi \left( 0 \right) \right\rangle$ 
is the ground state of the initial Hamiltonian 
$\mathcal{H}\left( 0 \right)$.
Figure 2 shows the energy as a function of the transverse field, where 
$J=1$, $J'=0.1$, $\Gamma_0 = 1.5$, and $N=6$.
It should be noted that there are two-fold evident degeneracy because of the up-down symmetry of spin.
In Fig. 2, we find that the ground state at $\Gamma = \Gamma_0$ is connected to that of $\Gamma=0$. 
Thus the adiabatic motion leads the system to the true ground state of $\Gamma=0$. 
In this sense, the success of quantum annealing is insured. 

Now we study properties of the quantum annealing. First, the $\tau$ dependence of
the probability $P_0$ to reach the ground state of $\mathcal{H}_0$.
Fig. 3 shows $P_0$ as a function of $\tau$.
There we find that the slower the speed of the sweep, the larger $P_0$. 
In the quantum mechanical system, we have several method to obtain the ground state 
such as the power method, Lanczos method {\it etc}. 
These methods are also possible processes to obtain the ground state. 
We will compare these methods as the annealing process and be reported elsewhere.
%Next, we consider the time dependence of the probability of staying 
%the ground state at each time $\left| G\left( t\right) \right\rangle$.
%Figure 4 shows $Q_0\left( t \right) = \left| \left\langle G(t) | \Phi (t) \right\rangle \right|^2$ 
%for various values of $\tau$.
%When we sweep the field slow enough, the state changes almost adiabatically.
%As the sweeping speed increases, the state is scattered from the ground state around $\Gamma=0.2$. 

%%%%%%%%%%%%%%%%%%%% Kondo in title, abstract and/or keywords %%%%%%%%%%%%%%%%
  \begin{figure}[h]
  \begin{center}
\includegraphics[scale=0.37]{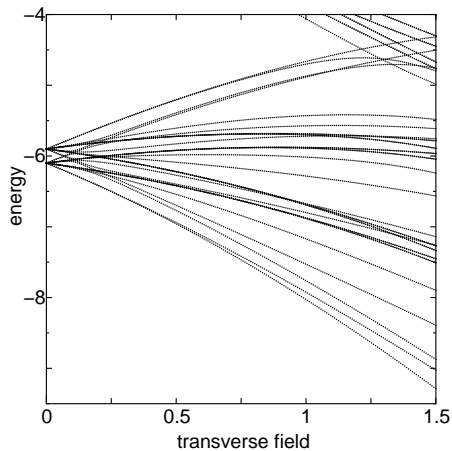}
\end{center}
\caption{
The energy diagram of decorated bond system ($N=6$) as a function of transverse field.
}
  	\label{fig-1}
  \end{figure}
%%%%%%%%%%%%%%%%%%%%%%%%%%%%%%%%%%%%%%%%%%%%%%%%%%%%%%%%%%%%%%%%%%%%%%%%%%%%%%

%%%%%%%%%%%%%%%%%%%% Kondo in title, abstract and/or keywords %%%%%%%%%%%%%%%%
%  \begin{figure}[t]
%  \begin{center}
%\includegraphics[scale=0.35]{tanaka-2-6-000434-f4.eps}
%\end{center}
%\caption{
%The probability of staying the ground state as a function of transverse field.
%}
%  	\label{fig-1}
%  \end{figure}
%%%%%%%%%%%%%%%%%%%%%%%%%%%%%%%%%%%%%%%%%%%%%%%%%%%%%%%%%%%%%%%%%%%%%%%%%%%%%%

\section{Conclusion}
Here we assume ferromagnetic interaction between $\sigma_1$ and $\sigma_2$.
thus the Fig. 1(a) gives the ground state. However if $\sigma_1$ and $\sigma_2$ are
antiparallel initially, the decorated spins denoted by rightward triangles align upwards and
the decorated spins denoted as the right half of the triangles remain disorder.
In order to change $\sigma_1$ and $\sigma_2$ antiparallel to parallel, 
all of the left half of the triangle must align, which is difficult.
This is called "entropically slowing down".
Here we treated an 'easy' example to find the ground state 
and to demonstrate the efficiency of the quantum annealing.
The decorated bond systems with larger number of $N$ ({\it e.g.} $N=20$) 
have been studied to research the temperature dependent structure \cite{Tanaka1}.
When $N$ increases the energy levels are much closer and the annealing becomes more 
difficult, which will be reported elsewhere \cite{Tanaka2}.

The present work is partially supported by Grand-in-Aid from the Ministry of Education, 
Culture, Sports, Science, and Technology, and also by NAREGI Nanoscience Project, Ministry of 
Education Culture, Sports, Science, and Technology, Japan.

%%%%%%%%%%%%%%%%%%%% Kondo in title, abstract and/or keywords %%%%%%%%%%%%%%%%
  \begin{figure}[h]
  \begin{center}
\includegraphics[scale=0.35]{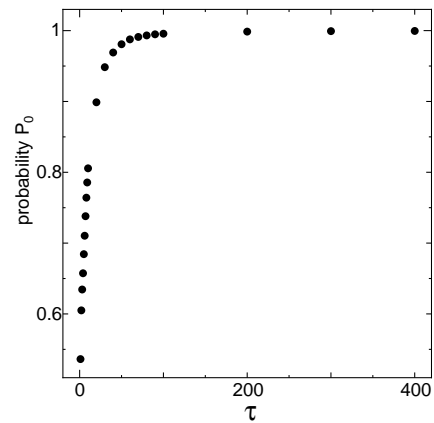}
\end{center}
\caption{ 
The probability of reaching the ground state of $\Gamma = 0$ 
as a function of $\tau$.
}
  	\label{fig-1}
  \end{figure}

\end{document}